\documentstyle[eqsecnum,preprint,aps,prd,epsf,rotate]{revtex}
\date{\today}
\draft
\tighten
\begin{document}
\def\sqr#1#2{{\vcenter{\hrule height.3pt
      \hbox{\vrule width.3pt height#2pt  \kern#1pt
         \vrule width.3pt}  \hrule height.3pt}}}
\def\square{\mathchoice{\sqr67\,}{\sqr67\,}\sqr{3}{3.5}\sqr{3}{3.5}}
\def\today{\ifcase\month\or
  January\or February\or March\or April\or May\or June\or July\or
  August\or September\or October\or November\or December\fi
  \space\number\day, \number\year}

\def\Bbb{\bf}

\preprint{DAMTP-1998-85}

\title{Inflation, Singular Instantons and Eleven Dimensional Cosmology}

\author{S.W. Hawking \thanks{email: S.W.Hawking@damtp.cam.ac.uk} and 
Harvey S. Reall \thanks{email: H.S.Reall@damtp.cam.ac.uk}}

\address {\qquad \\ DAMTP\\
Silver Street\\
Cambridge, CB3 9EW, UK
}

\maketitle

\begin{abstract}

We investigate cosmological solutions of eleven dimensional
supergravity compactified on a squashed seven manifold. The effective
action for the four dimensional theory contains scalar fields
describing the size and squashing of the compactifying space. The
potential for these fields consists of a sum of exponential terms. At
early times only one such term is expected to dominate. The condition
for an exponential potential to admit inflationary solutions is
derived and it is shown that inflation is not possible in our model.
The criterion for an exponential potential to admit a
Hawking-Turok instanton is also derived. It is shown that the
instanton remains singular in eleven dimensions.

\end{abstract}


\section{Introduction}

Until recently it was thought that slow-roll inflation always gives rise to a
flat universe. This assumption was proved incorrect in
\cite{bucher,yamamoto} (building on the earlier work of \cite{coleman} and
\cite{gott} on bubble nucleation and `old' inflation) where it was 
demonstrated that an open universe can arise
after quantum tunneling of a scalar field initially trapped in a false
vacuum. (One can also obtain open inflation in two field models
\cite{linde1,linde2}). 
However such models of open inflation appear rather contrived
owing to the special form that the scalar potential must be assumed to
take. They also do not address the problem of the initial conditions
for the universe i.e. no explanation is given of how the scalar field
became trapped in the false vacuum. These two objections were
confronted in \cite{ht1} within the framework of the
`No Boundary Proposal' \cite{hartle}. It was described there how
an open universe could be created without assuming any special form
for the potential. The approach was to construct an instanton
(i.e. a solution to the Euclidean field equations) and analytically
continue to Lorentzian signature. The novel feature of the
instanton is that it is singular although the
singularity is sufficiently mild for the instanton to possess a finite
action. Several objections have been raised against the use of such
instantons, the most serious of which is Vilenkin's argument
\cite{vilenkin} that if such instantons are allowed then flat space
should be unstable to the nucleation of singular bubbles. Another
objection is that the singularity can be viewed as a boundary of the
instanton (there is a finite contribution to the action from the
boundary \cite{vilenkin}) which is unacceptable according to the no boundary
proposal.

There have been three different approaches to dealing with the problems
raised by a singular instanton. The first is to regularize the singularity with
matter in the form of a membrane \cite{garriga1,bousso}. 
An alternative approach
\cite{ht2} is to analytically continue the instanton across a
deformed surface that does not include the singularity. The problem
with this is that the surface does not have vanishing second
fundamental form which means that one obtains a region of spacetime which
does not have purely Lorentzian signature. It was pointed out
that this region is not in the open universe so it may not have
observable consequences. The third approach, due to 
Garriga \cite{garriga2}, is to construct
a four dimensional singular instanton from a higher dimensional
non-singular one. This approach is particularly appealing because
$M$-theory is eleven dimensional. Garriga gives a non-singular five
dimensional instanton that reduces to Vilenkin's in four dimensions but with a
cut-off to the scale of bubble nucleation that makes the decay rate of
flat space unobservably small. He also gives a
five dimensional solution with cosmological constant that reduces to a
four dimensional instanton of Hawking-Turok type. (Garriga's 
five dimensional instantons are just Euclidean
Schwarzschild and the five sphere respectively). One purpose of this
paper is to examine whether it is possible to
obtain Hawking-Turok instantons in four dimensions from non-singular
instantons of eleven dimensional supergravity, the low energy
limit of $M$-theory.

Our second aim is to investigate whether solutions of eleven
dimensional supergravity corresponding to four dimensional inflating
universes exist. Since inflation is now widely accepted as the
standard explanation of several cosmological problems (see
e.g. \cite{kolb}), one would expect the existence of inflationary
solutions of $M$-theory if it is indeed the correct theory of
everything. However compactifications of $D=11$ supergravity usually
give a $negative$ cosmological constant (see \cite{duff}) which is
precisely the opposite of what we need for inflation. The reason for
this is that if the compactifying space has positive curvature then
the field equations imply that our space has negative curvature. This
suggests that a way around the problem may be to look for solutions
with the seven dimensional compactifying space $M_7$ negatively curved
at early times but positively curved at late times. We do this by
taking $M_7$ to be a coset space and squashing it (the meaning of
squashing is explained below), treating the squashing parameters as
dynamical scalar fields.

Upon reduction to four dimensions we obtain a model with scalar fields
evolving according to a potential consisting of a sum of exponential
terms. At early times only one term in the sum is expected to be
significant. Cosmological solutions involving scalar fields with
exponential potentials have been investigated by several
workers. Lucchin and Matarrese \cite{lucchin} showed that power-law
inflation can result from such potentials. This was further
investigated by Barrow \cite{barrow} who gave an exact scaling
solution to the equations of motion which was subsequently generalised
by Liddle \cite{liddle}. Halliwell \cite{halliwell} has conducted a
phase-plane analysis of the equations of motion resulting from an
exponential potential. Wetterich has derived scaling solutions for
cosmologies with the scalar field coupled to other matter \cite{wetterich}.
For the single scalar case we have found a first integral of the 
equations of motion and give an exact expression for the number 
of inflationary efoldings. It is found that a significant inflationary
period only results from solutions that approach the scaling
solutions at late times.
The results are generalized to the multi-scalar case. We 
have analysed the behaviour of scalars with an 
exponential potential near the singularity of the 
instanton and give a criterion for the singularity to be
integrable. (This was discussed in \cite{copeland} but the
analysis was incomplete).

Applying the results on exponential potentials to our model from
eleven dimensions yields the disappointing result that the potential
is too steep for inflation to occur. We
find that unlike in Garriga's models the instanton is singular in
eleven dimensions. The reason
for this is that Garriga's potential comes from a five dimensional
cosmological constant whereas ours comes from the Ricci scalar of the
compactifying space and has too steep a dependence on the scalar field
that measures the size of the internal space (i.e. its `breathing'
mode). It is this same
dependence that rules out inflationary behaviour which leads us to
speculate that if one could fix the size of the internal space then a
solution with more appealing properties might be found. 

As this paper was nearing completion we received a paper by Bremer
{\it et al} \cite{bremer} which has some overlap with our work. They
also consider cosmological solutions with dynamical squashing in
various dimensions. Their $S^7$ example is not the same as ours: they
obtain squashed metrics on $S^7$ by viewing it as a $U(1)$ bundle over
$CP^3$ and squashing corresponds to varying the length of the $U(1)$
fibres whereas we treat $S^7$ as a $S^3$ bundle over $S^4$ and
squashing corresponds to varying the size of the $S^3$ fibres. Our
methods are applicable to any squashed coset space (although we always
use the Freund-Rubin ansatz \cite{freund}). Integrability
of the instanton singularity is not discussed in \cite{bremer} (indeed
the examples discussed there all appear to be non-integrable) and
neither is the condition for inflation. (In the conclusions
section of \cite{bremer} it is stated that the instanton solutions can
be continued to give open inflationary universes. This is not the
case: the potentials are too steep to yield a significant inflationary
period). 

\newpage

\section{Eleven dimensional supergravity}

The action for the bosonic sector of $D=11$ supergravity is \cite{duff}
\begin{eqnarray}
 \label{eqn:action}
 \hat{S} = \int
d^{11}x\sqrt{-\hat{g}}\left(\frac{1}{2\hat{\kappa}^2}\hat{R}\right.&&-\left.\frac{1}{48}\hat{F}_{MNPQ}\hat{F}^{MNPQ}+{}\right.
\nonumber\\ & & \left.{} +\frac{2}{(12)^4}\frac{1}{\sqrt{-\hat{g}}}\epsilon^{M_1\ldots
M_{11}}\hat{F}_{M_1\ldots M_4}\hat{F}_{M_5\ldots M_8}\hat{A}_{M_9\ldots M_{11}}\right)+S_{boundary}.
\end{eqnarray}
Hats will be used to distinguish eleven dimensional quantities from
four dimensional ones. Upper case Roman letters will be used for eleven 
dimensional indices, and lower case Greek letters for four dimensional ones. 
$\hat{\kappa}^2=8\pi\hat{G}$ is the eleven dimensional Planck
scale. We will use a positive signature metric and a curvature
convention such that a sphere has positive Ricci
scalar. $\epsilon^{M_1\ldots M_{11}}$ is the alternating tensor
density. The four form 
$\hat{F}_{MNPQ}$ is related to its three form potential $\hat{A}_{MNP}$ by
\begin{equation}
 \label{eqn:forms}
{\hat{F}_{MNPQ}=4\partial_{[M}\hat{A}_{NPQ]}}
\end{equation}
where square brackets denote antisymmetrization. 

$S_{boundary}$ is a sum of boundary terms which are essential
in quantum cosmology:
\begin{equation}
 {S_{boundary}=B_1+B_2,}
\end{equation}
where $B_1$ is the Gibbons-Hawking boundary term \cite{gibbons} and
$B_2$ is needed because we want to consider the Hartle-Hawking
wavefunction \cite{hartle} as a function of the four-form on the
boundary so it is the variation of the four form that should vanish 
on the boundary, not that of the three form. See \cite{ht2} for a 
discussion of this point. We shall only consider solutions with
vanishing Chern-Simons term, for which
\begin{equation}
 {B_2=\frac{1}{6}\int
 d^{11}x\partial_M\left(\sqrt{-\hat{g}}\hat{F}^{MNPQ}\hat{A}_{NPQ}\right).}
\end{equation}
The equations of motion
following from the action \ref{eqn:action} are:
\begin{equation}
 \label{eqn:einsteineq}
{\hat{R}_{MN}=\frac{\hat{\kappa}^2}{6}\left(\hat{F}_{MPQR}\hat{F}_N^{PQR}-\frac{1}{12}\hat{F}_{PQRS}\hat{F}^{PQRS}\hat{g}_{MN}\right),}
\end{equation}
\begin{equation}
 \label{eqn:feq}
{\partial_M\left(\sqrt{-\hat{g}}\hat{F}^{MNPQ}\right)=-\frac{1}{576}\epsilon^{NPQM_1\ldots
M_8}\hat{F}_{M_1\ldots M_4}\hat{F}_{M_5\ldots M_8}.}
\end{equation}
If $F\wedge F\wedge n$ vanishes on the boundary (where $n$ is the
1-form normal to the boundary) then the action is gauge invariant and
the second boundary term is
\begin{equation}
 {B_2=\frac{1}{24}\int d^{11}x\sqrt{-\hat{g}}\hat{F}^{MNPQ}\hat{F}_{MNPQ}.}
\end{equation}

\section{Squashed manifolds}

Given a Lie group $G$, the manifolds admitting a
transitive action of $G$ can be viewed as coset spaces $G/H$ where $H$
is the isotropy subgroup. We are interested in 
the most general $G$-invariant metric on such a manifold (i.e. the most general
metric for which the left action of $G$ yields a group of
isometries). If $G/H$ is isotropy irreducible (see \cite{duff})
then there is a unique (up to scale) such metric which is actually an Einstein
metric. For example if $G=SO(8)$ and $H=SO(7)$ then the unique
$G$-invariant metric on $G/H$ is the round metric on $S^7$. 
If the coset space is not isotropy irreducible then the general
$G$-invariant metric contains arbitrary parameters. This is what is
meant by squashing. An example is $G=SO(5)$ and $H$ the $SO(3)$
subgroup such that $G/H$ has $S^7$
topology. The most general $G$-invariant metric contains seven
arbitrary parameters and there are $\it{two}$ $G$-invariant Einstein
metrics.

It is discussed in \cite{castellani} how one can squash a coset space
by rescaling the vielbein i.e. $e^a\rightarrow \lambda_a e^a$ (no
summation). A criterion is given for deciding if a
particular rescaling will preserve the isometry group of the
metric. This is the most general kind of deformation that
preserves the $G$-invariant metric on $G/H$ (see \cite{camporesi} for
a review).

\section{Dimensional reduction with dynamical squashing}

Our metric ansatz is
\begin{equation}
 \label{eqn:ansatz}
{d\hat{s}^2=e^{2B(x)}g_{\mu\nu}(x)dx^{\mu}dx^{\nu}+g_{mn}(x,y)dy^mdy^n,}
\end{equation}
where $x^{\mu}$ and $y^m$ are coordinates on the four and seven dimensional 
manifolds with metrics $g_{\mu\nu}$ and $g_{mn}$ respectively. We shall 
choose the
field $B(x)$ so that the reduced action is in the Einstein
frame. The siebenbein on the internal manifold is assumed to be
\begin{equation}
 {e^a_m(x,y)=e^{A_a(x)}\overline{e}^a_m(y)\qquad \textrm{(no
 summation),}}
\end{equation}
where $\overline{g}_{mn}(y)=\sum_a
\overline{e}^a_m(y)\overline{e}^a_n(y)$ is the unsquashed metric.
The squashing is described by the seven scalar fields $A_a(x)$. Note
that for squashing in the sense described above (i.e. preserving
the isometry group) these scalar fields will not be independent.

In the following discussion we shall not specify a particular squashed
coset for $M_7$. The choice is not arbitrary: the eleven dimensional
field equations have to be satisfied. We shall assume that a suitable
coset has been found but our conclusions will be independent of the
details of the internal space. 
As an example we shall consider $S^7$ as a $SO(5)/SO(3)$ coset with a
two parameter family of metrics (i.e. only two of the $A_a$ are
independent), one parameter being the size and the other the
squashing. This choice does satisfy the $D=11$ field
equations. 

$B(x)$ is calculated by observing
\begin{equation}
\label{eqn:bsoln}
 {\sqrt{-\hat{g}}\hat{R}=\sqrt{-g}\sqrt{\overline{g}_7}e^{\sum
 A_a(x)}e^{2B(x)}\left(R+\ldots\right),}
\end{equation}
so after integrating over $y^m$ the reduced action will be in the
Einstein frame provided that
\begin{equation}
 {B(x)=-\frac{1}{2}\sum_a A_a(x).}
\end{equation}
In the Einstein frame the components of the eleven dimensional Ricci
tensor are (see appendix):
\begin{equation}
 {\hat{R}_{\mu\nu}=R_{\mu\nu}+\frac{1}{2}\nabla^2\left(\sum_a{A_a}\right)g_{\mu\nu}-\frac{3}{2}\sum_a A^a_{,\mu}A^a_{,\nu}-\frac{1}{2}\sum_{a\neq b}A^a_{,\mu}A^b_{,\nu},}
\end{equation}
\begin{equation}
 {\hat{R}_{ab}=R_{ab}\left[M_7\right]-e^{\sum
 A_a}\textrm{diag}\left(\nabla^2 A_a\right).}
\end{equation}
The four dimensional part has been written with curved indices and the
seven dimensional part with tangent space indices for notational
clarity.

We shall use the Freund-Rubin ansatz \cite{freund} for the four form i.e.
\begin{equation}
 \label{eqn:freund}
 {\hat{F}_{\mu\nu\rho\sigma}(x,y)=F_{\mu\nu\rho\sigma}(x)\qquad\textrm{other
components vanish.}}
\end{equation}
(Some other ans\"atze for the four form were considered in \cite{bremer}).

One can substitute these ans\"atze into the field equations to obtain 
equations of motion for the effective four dimensional
theory. Alternatively one can obtain the same equations by varying the
reduced action obtained by substituting into \ref{eqn:action}:
\begin{eqnarray}
 S  = \int d^4x\sqrt{-g}
 \left(\frac{1}{2\kappa^2}R\right.&-&\frac{1}{4\kappa^2}\sum_{a,b}(\partial
 A_a)M_{ab}(\partial A_b)+{} \nonumber \\ {}&+&\left.\frac{1}{2\kappa^2}e^{-\sum
 A_a}R\left[M_7\right]-\frac{1}{48}e^{3\sum A_a}F_{\mu\nu\rho\sigma}F^{\mu\nu\rho\sigma}\right)+B,
\end{eqnarray}
where $B$ is a boundary term and indices are raised with $g^{\mu\nu}$. 
The integration over $y^m$ gives a volume factor $V_7=\int
d^7y\sqrt{\overline{g}_7}$ which is absorbed into the
definition of the four dimensional Planck scale:
$\kappa^2=\hat{\kappa}^2 V_7$. A factor of $\sqrt{V_7}$ has also been
absorbed into $F_{\mu\nu\rho\sigma}$. The matrix $M_{ab}$ has threes on
its diagonal and ones everywhere else. $R\left[M_7\right]$ is the
(seven dimensional) Ricci scalar of the internal space computed treating
$A_a(x)$ as constant parameters.

Note that there is no guarantee that solutions of the four dimensional
equations of motion obtained from this action are solutions of the
eleven dimensional field equations. This is because we have not
considered the field equation associated with the seven internal
dimensions. However if one chooses a coset such that this field
equation can be satisfied then the resulting equations of motion will
be the same as those obtained from the reduced action. In our $S^7$
example the Ricci tensor of the internal space (see appendix B) splits
into two independent diagonal parts. This will give two independent
field equations. In order to satisfy them (at least for non-constant
scalar fields), we must include at least
two degrees of freedom in the metric on $S^7$. So in addition to
squashing $S^7$ we allow its size to vary. This is achieved by
multiplying its metric by an overall conformal factor
$e^{2C(x)}$. Then the scalars $A_a$ are given by
\begin{equation}
 {A_1=A_2=A_3=A_4=C,\qquad A_5=A_6=A_7=A+C,}
\end{equation}
where $e^{A}$ is the squashing parameter defined in appendix B. The
two field equations coming from the internal space give the equations
of motion for $A$ and $C$. The same equations of motion can be obtained
from the four dimensional reduced action. 

Returning to the general case, the kinetic term can be diagonalised by defining
\begin{equation}
 {\phi_k=\frac{1}{\kappa\sqrt{k(k+1)}}\left(\sum_{j=1}^k
 A_j-kA_{k+1}\right)\qquad k=1\ldots 6,}
\end{equation}
\begin{equation}
 {\psi=\frac{3}{\kappa\sqrt{14}}\sum_{j=1}^{7}A_j.}
\end{equation}
If the scalar fields $A_a$ are not linearly independent then the fields
$\phi_k$ will not be independent and the kinetic terms will still not
be correctly normalised. This occurs in our squashed $S^7$ example:
\begin{eqnarray}
 \phi_1=\phi_2=\phi_3=0, & &
 \sqrt{4.5}\phi_4=\sqrt{5.6}\phi_5=\sqrt{6.7}\phi_6=
 -\frac{4A}{\kappa},\nonumber \\ \psi & = &\frac{3}{\kappa\sqrt{14}}(3A+7C).
\end{eqnarray}
Since $\phi_4$, $\phi_5$ and $\phi_6$ are not independent we define
$\phi=\sqrt{\frac{12}{7}}\frac{A}{\kappa}$ so that
\begin{equation}
 {\frac{1}{2}(\partial\phi_4)^2+\frac{1}{2}(\partial\phi_5)^2+\frac{1}{2}(\partial\phi_6)^2=\frac{1}{2}(\partial\phi)^2,}
\end{equation}
so now the scalar fields $A$ and $C$ have been replaced by $\phi$ and
$\psi$ with diagonal kinetic terms. 

Note that a scaling of the internal manifold $A_a(x)\rightarrow
A_a(x)+C(x)$ only affects $\psi$, which measures its size. In general
one must allow the size to vary in order to satisfy the $D=11$ field
equations (i.e. one could not impose $\psi=\textrm{constant}$ except
in special cases corresponding to static solutions) hence $\psi$ and
$\phi_k$ will be independent. Thus the kinetic term for $\psi$
$\it{is}$ correctly normalized and $\psi$ will not need
rescaling. It is this that will allow us to draw general conclusions
later on about the possibility of inflation or higher dimensional
non-singular instantons in our model.

The inverse transformation relating $A_j$ to $\phi_k$ and $\psi$ is
\begin{equation}
\label{eqn:inverse}
 {A_j=\kappa\left(-\sqrt{\frac{j-1}{j}}\phi_{j-1}+\sum_{k=j}^6\frac{1}{\sqrt{k(k+1)}}\phi_k+\frac{\sqrt{14}}{21}\psi\right).}
\end{equation}
Substituting into the reduced action gives
\begin{eqnarray}
 S & = & \int
 d^4x\sqrt{-g}\left(\frac{1}{2\kappa^2}R-\sum_{k=1}^{6}\frac{1}{2}(\partial \phi_k)^2-\frac{1}{2}(\partial \psi)^2 -W(\phi_k,\psi)-{}\right. \nonumber\\ & & \left.{}-\frac{1}{48}e^{\kappa\sqrt{14}\psi}F_{\mu\nu\rho\sigma}F^{\mu\nu\rho\sigma}\right)+B
\end{eqnarray}
where the scalar potential is
\begin{equation}
 {W(\phi_k,\psi)=-\frac{1}{2\kappa^2}e^{-\frac{\sqrt{14}}{3}\kappa\psi}R\left[M_7\right].}
\end{equation}

The equations of motion following from this action are:
\begin{eqnarray}
 \label{eqn:einsteineq2}
 R_{\mu\nu}=2\kappa^2\left[\sum_k\frac{1}{2}\partial_{\mu}\phi_k\partial_{\nu}\phi_k\right. &+&\frac{1}{2}\partial_{\mu}\psi\partial_{\nu}\psi+\frac{1}{2}Wg_{\mu\nu} + {} \nonumber \\ {} & + &\left.\frac{1}{12}e^{\kappa\sqrt{14}\psi}\left(F_{\mu\rho\sigma\tau}F_{\nu}^{\rho\sigma\tau}-\frac{3}{8}F_{\lambda\rho\sigma\tau}F^{\lambda\rho\sigma\tau}g_{\mu\nu}\right)\right],
\end{eqnarray}
\begin{equation}
 \label{eqn:phieq}
 {\nabla^2\phi_k=\frac{\partial W}{\partial\phi_k},}
\end{equation}
\begin{equation}
 \label{eqn:psieq}
 {\nabla^2\psi=\frac{\partial W}{\partial \psi}+\frac{\kappa\sqrt{14}}{48}e^{\kappa\sqrt{14}\psi}F_{\mu\nu\rho\sigma}F^{\mu\nu\rho\sigma},}
\end{equation}
\begin{equation}
 \label{eqn:Feq2}
 {\partial_{\mu}(\sqrt{-g}e^{\kappa\sqrt{14}\psi}F^{\mu\nu\rho\sigma})=0.}
\end{equation}
Note that the final equation is obtained by varying $A_{\mu\nu\rho}$. This
equation has the unique solution
\begin{equation}
 \label{eqn:Fsoln}
 {F_{\mu\nu\rho\sigma}=\sqrt{-g}\epsilon_{\mu\nu\rho\sigma}e^{-\kappa\sqrt{14}\psi}F,}
\end{equation}
for some constant $F$. If we substitute this solution into \ref{eqn:psieq}
then we get
\begin{equation}
 \label{eqn:psieq2}
 {\nabla^2\psi=\frac{\partial V}{\partial \psi},}
\end{equation}
where
\begin{equation}
 \label{eqn:Wdef}
 {V(\phi_k,\psi)=W(\phi_k,\psi)+\frac{1}{2}F^2e^{-\kappa\sqrt{14}\psi}}
\end{equation}
is the effective potential that determines the evolution of the field $\psi$.
Note that we can replace $W$ by $V$ in the field equation for $\phi_k$
and substitute the solution for $F_{\mu\nu\rho\sigma}$ into
\ref{eqn:einsteineq2} to yield
\begin{equation}
 {R_{\mu\nu}=2\kappa^2\left(\sum_k\frac{1}{2}(\partial_{\mu}\phi_k)(\partial_{\nu}\phi_k)+\frac{1}{2}(\partial_{\mu}\psi)(\partial_{\nu}\psi)+\frac{1}{2}V g_{\mu\nu}\right),}
\end{equation}
so now $V$ occurs in all of the equations of motion and one can forget
about $W$. 

For our squashed $S^7$ example, the potential is
\begin{equation}
 {V(\phi,\psi)=-\frac{1}{2\kappa^2}e^{-\frac{3}{7}\sqrt{14}\kappa\psi}\left(\frac{3}{2}e^{-\frac{4}{21}\sqrt{21}\kappa\phi}+12e^{\frac{1}{7}\sqrt{21}\kappa\phi}-3e^{\frac{10}{21}\sqrt{21}\kappa\phi}\right)+\frac{1}{2}F^2 e^{-\sqrt{14}\kappa\psi}.}
\end{equation}
Plotted as a function of $\phi$ this tends to $\pm\infty$ as
$\phi\rightarrow\pm\infty$. There is a local minimum at $\phi=0$
corresponding to the round metric on the $S^7$. There is also a local
maximum at a negative value of $\phi$ corresponding to the squashed
Einstein metric on $S^7$. The qualitative behaviour as $\psi$ varies
depends on the value of $\phi$. There is a positive constant $\phi_0$
such that at $\phi=\phi_0$ the $F$-independent part of the potential
vanishes. For $\phi>\phi_0$, $V$ is a monotonically decreasing
function of $\psi$ tending to $+\infty$ as $\psi\rightarrow -\infty$
and to $0$ as $\psi\rightarrow +\infty$. For $\phi<\phi_0$ (which
includes the two Einstein metrics), the asymptotic behaviour is
similar but there is a local minimum at some value of $\psi$
corresponding to a negative value of $V$. Hence there exist static
solutions of $D=11$ supergravity with $\psi$ sitting at this minimum
and $\phi$ corresponding to either the round or the squashed Einstein
metric. These have been extensively discussed from the point of view of
Kaluza-Klein theory \cite{duff}. We are interested in solutions with
a positive potential at early times in the hope that these may exhibit
inflationary behaviour. Such solutions start with $\phi$ large and
positive, corresponding to a negatively curved metric on $S^7$. One
would expect solutions to exist in which $\phi$ rolls down to the
local minimum so the solution settles into the Freund-Rubin solution 
$AdS_4\times S^7$ (with a round metric) at late times. Note that this
solution appears unstable because $\phi$ can tunnel past the local maximum
and roll off to $-\infty$. However Breitenlohner and Freedman
\cite{breit} have shown how boundary conditions at infinity can
stabilise $AdS$, at least against small perturbations, so one would
expect a similar argument to be valid here. 

A second example that we have considered involves taking the
compactifying space to be $S^1\times S^3\times S^3$. 
$S^3$ is group manifold so one can squash all 
three directions independently \cite{duff}. Thus this $M_7$ can be
squashed with all seven $A_a$ independent. The $D=11$ field equations
can be satisfied with this $M_7$. The Ricci scalar of $S^3$ with
squashing described by $A_5$, $A_6$ and $A_7$ is
\begin{equation}
\label{eqn:s3ricci}
 {R=e^{-2A_5}+e^{-2A_6}+e^{-2A_7}-\frac{1}{2}\left(e^{2(A_5-A_6-A_7)}+e^{2(A_6-A_7-A_5)}+e^{2(A_7-A_5-A_6)}\right).}
\end{equation}
Thus in order to get a negatively curved $S^3$ the second group of
terms must dominate the first. Note that there is no static
Freund-Rubin solution in this case because $S^1\times S^3\times S^3$
cannot be given an Einstein metric. Thus the potential $V$ does not
have any extrema. 

\section{Condition for inflation}

We shall now seek a solution of the field equations derived above that
describes a four dimensional universe. Spatial homogeneity and
isotropy imply that the metric must take the form
\begin{equation}
 \label{eqn:cosmetric}
 {ds^2=-dt^2+a(t)^2ds_3^2}
\end{equation}
where $ds_3^2$ is the line element of a three-space of constant curvature.
Substituting this into \ref{eqn:einsteineq2} yields the equations
\begin{equation}
 \label{eqn:einsteinconstraint}
 {\left(\frac{\dot{a}}{a}\right)^2=\frac{\kappa^2}{3}\left(\sum_k\frac{1}{2}\dot{\phi_k}^2+\frac{1}{2}\dot{\psi}^2+V\right)-\frac{k}{a^2},}
\end{equation}
\begin{equation}
 \label{eqn:atwodots}
 {\ddot{a}=-\frac{\kappa^2}{3}a\left(\sum_k\dot{\phi_k}^2+\dot{\psi}^2-V\right).}
\end{equation}
$k$ is the sign of the curvature of the spatial sections. 

Inflation is defined by $\ddot{a}>0$ so \ref{eqn:atwodots} implies
that for inflation we need 
\begin{equation}
 \label{eqn:inflationcondition}
 {V>\sum_k\dot{\phi_k}^2+\dot{\psi}^2.}
\end{equation}

Since the potentials we have obtained consist of a sum of exponential
terms with no extrema of $\psi$ at positive values of the potential, one would
expect that if inflation occurs then it would do so at a large value
of the potential where typically only one exponential is
significant. For simplicity we shall consider a single scalar model 
\begin{equation}
 {V(\phi)=V_0e^{\alpha\kappa\phi}.}
\end{equation}
The equation of motion for the scalar field is
\begin{equation}
 {-\ddot{\phi}-3H\dot{\phi}=\frac{d V}{d\phi}}
\end{equation}
where the Hubble parameter is $H(t)=\frac{\dot{a}}{a}$. We shall
assume that $\alpha>0$, which can always be achieved by reversing the
signs of $\alpha$ and $\phi$. 

For inflation we need $V$ to be larger than the scalar kinetic term
and curvature term (if $k\neq 0$) so one would 
expect the Hubble parameter to behave like
$e^{\frac{1}{2}\alpha\phi}$. (We have set $\kappa=1$). After 
substituting this into the equation
of motion for the scalar field it is natural to seek a solution of the
form $\dot{\phi}\propto e^{\frac{1}{2}\alpha\phi}$. With this in
mind, define a new variable by
\begin{equation}
 {\Phi = -\dot{\phi}e^{-\frac{1}{2}\alpha\phi}.}
\end{equation}
If one replaces $\ddot{\phi}$ by
$\frac{d}{d\phi}\left(\frac{1}{2}\dot{\phi}^2\right)$, eliminates
$\dot{\phi}$ in favour of $\Phi$ and neglects the curvature term (this
is the only approximation that we shall make) then one obtains
\begin{equation}
 {\frac{1}{2}\frac{d}{d\phi}\Phi^2+\frac{\alpha}{2}\Phi^2-\sqrt{3\left(\frac{1}{2}\Phi^2+V_0\right)}\Phi+\alpha V_0=0.}
\end{equation}
Now define $x$ by $\Phi=\sqrt{2V_0} \sinh x$ to give
\begin{equation}
\label{eqn:xeqn}
 {\frac{d x}{d\phi}=\frac{1}{2}\left(\sqrt{6}-\alpha\coth
 x\right).}
\end{equation}
It is obvious that there is a solution with
$x=\textrm{constant}$ when $\alpha<\sqrt{6}$. This is the solution
obtained previously by Barrow\cite{barrow} and Liddle\cite{liddle}. 
However we can investigate the general solution by using the
change of variable $y=e^{2x}$ to give
\begin{equation}
\label{eqn:phisoln}
 {e^{\frac{1}{2}\alpha
 (\phi-\phi_*)}=F(y)\equiv y^{\frac{\alpha}{2(\sqrt{6}+\alpha)}}\left|y-\frac{\sqrt{6}+\alpha}{\sqrt{6}-\alpha}\right|^{\frac{\alpha^2}{6-\alpha^2}}.}
\end{equation}
$\phi_*$ is a constant of integration. If we are interested in real
solutions obtained by analytical continuation from a Euclidean
instanton then we must impose the initial conditions
\begin{equation}
 {\phi=\phi_0,\qquad \dot{\phi}=0 \Rightarrow y=1.}
\end{equation}
In the model of open inflation described in \cite{ht1}, one
analytically continues the instanton to an open universe at a point
where the scale factor vanishes. This means that initially it is not a
good approximation to neglect the curvature term in the Einstein
constraint equation as we have done here. However if there is a
significant inflationary period then the curvature term will rapidly
become negligible. So, strictly speaking, our analysis is only
applicable after this term has become negligible by which time the
above boundary conditions will not hold (since then $\dot{\phi}<0$ so
$y>1$). However, as we shall show, the condition for a significant
period of inflation is not sensitive to the initial value of
$\dot{\phi}$ so we shall take the above value for simplicity. Of
course our results are exact for flat ($k=0$) universes.

With these boundary conditions, \ref{eqn:phisoln} becomes
\begin{equation}
 {e^{\frac{1}{2}\alpha(\phi-\phi_0)}=\frac{F(y)}{F(1)}.}
\end{equation}
The condition for inflation is $\dot{\phi}^2<V$, or equivalently,
$\sinh^2 x<\frac{1}{2}$. This is satisfied if, and only if,
\begin{equation}
 {2-\sqrt{3}<y<2+\sqrt{3}.}
\end{equation}
Since we are starting with $\dot{\phi}=0$ we will always get \emph{some}
inflation. How much we get depends on $F(y)$, the qualitative
behaviour of which depends on the magnitude of $\alpha$. There are two
cases to consider: \emph{i}) $\alpha^2<6$ and \emph{ii}) $\alpha^2>6$.
In the first case $F(y)$ has zeros at $y=0$ and
$y=y_0\equiv\frac{\sqrt{6}+\alpha}{\sqrt{6}-\alpha}$ and a local maximum at
$y=1$. For large $y$, $F(y)$ tends to infinity as a power of $y$. The
solution $x=\textrm{constant}$ corresponds to the second zero of
$F(y)$ (but this solution is incompatible with the boundary condition 
$\dot{\phi}=0$ since it corresponds to an eternally inflating universe).

If $\alpha^2<2$ then the second zero of $F(y)$ lies within the range of
values of $y$ corresponding to inflation. This implies that the
solution will inflate all the way to $F(y)=0$ i.e. to
$\phi=-\infty$. For larger $\alpha$ inflation will stop before
$F(y)=0$ is reached. In case \emph{ii}) the only zero of $F(y)$ is at
$y=0$. Once again $F(y)$ has a maximum at $y=1$, beyond which it
decreases monotonically to zero as $y\rightarrow\infty$. 

We have succeeded in finding a first integral for the scalar field
equation of motion. This relates $\phi$ and $\dot{\phi}$
implicitly. It does not seem possible to integrate this again to find
an explicit solution for $\phi(t)$ but this is not necessary in order
to calculate the number of inflationary efoldings $N$, defined by
\begin{equation}
 {N=\int_0^{t_{max}}H(t)dt,}
\end{equation}
where $t_{max}$ is the (comoving) time at which inflation stops. One
can now substitute the expression for $H(t)$ in terms of $\Phi$ and
$\phi$ and then substitute for $\Phi$ and $\phi$ in terms of $y$ using
the definition of $y$ and \ref{eqn:phisoln}. To transform the integral
over $t$ into an integral over $y$ one needs to know $\frac{d y}{d t}$
which is obtained by differentiating \ref{eqn:phisoln} with respect to
$t$. $y$ runs from $1$ (when $\dot{\phi}=0$) to $2+\sqrt{3}$ (end of 
inflation). This gives
\begin{equation}
\label{eqn:nint} 
{N=\frac{2}{\alpha\sqrt{3}}\int_1^{2+\sqrt{3}}\left(\frac{y+1}{y-1}\right)\left(\frac{-F'(y)}{F(y)}\right)dy,}
\end{equation}
which is infinite if $\alpha<\sqrt{2}$ and otherwise evaluates to
\begin{equation}
 {N=\frac{\sqrt{6}}{3(\sqrt{6}+\alpha)}\left[\frac{1}{2}\log
 (2+\sqrt{3})+ \frac{\sqrt{6}}{\sqrt{6}-\alpha}\log\left(
 1+\frac{\sqrt{6}-\alpha}{\sqrt{3}(\alpha-\sqrt{2})} \right) \right].}
\end{equation}
This is small unless $\alpha$ is exponentially close to $\sqrt{2}$. We
can conclude that an exponential potential can only give a significant
inflationary period if $\alpha\leq\sqrt{2}$. Note that the result is
independent of $\phi_0$ in contrast with the result for power law
potentials. As mentioned above, the initial value of $\dot{\phi}$ does
not significantly affect the amount of inflation as can be verified by
changing the lower limit of integration in \ref{eqn:nint}.

It is easy to calculate the asymptotic behaviour of the solutions
found above. $\phi\rightarrow -\infty$ at late times so
$F(y)\rightarrow 0$. Hence $y\rightarrow y_0$ in case \emph{i}) and
$y\rightarrow\infty$ in case \emph{ii}). In the first case one has
$\Phi\rightarrow\Phi_0=\textrm{constant}$ so using the definition of
$\Phi$ one obtains the solution for $\phi(t)$. The kinetic and
potential energy densities and the scale factor have the following
asymptotic behaviour
\begin{equation}
 {\frac{1}{2}\dot{\phi}^2=\frac{2}{\alpha^2(t-t_0)^2},}
\end{equation}
\begin{equation}
 {V=V_0 e^{\alpha\phi} =
 \frac{6-\alpha^2}{\alpha^4}\frac{2}{(t-t_0)^2},}
\end{equation}
\begin{equation}
 {a=a_0 (t-t_0)^{\frac{2}{\alpha^2}}.}
\end{equation}
It is clear from these expression that it is only consistent to
neglect the curvature term in the Einstein constraint equation at large times
if $\alpha^2<2$, otherwise these results are restricted to flat
($k=0$) cosmologies. 

In case \emph{ii}) $y\rightarrow \infty$ implies $x\rightarrow
\infty$. Substituting this into \ref{eqn:xeqn} and solving gives the
following
\begin{equation}
 {\frac{1}{2}\dot{\phi}^2=\frac{1}{3(t-t_0)^2},}
\end{equation}
\begin{equation}
 {V=V_0 e^{\alpha\phi} =
 \frac{1}{(t-t_0)^{\frac{2\alpha}{\sqrt{6}}}},}
\end{equation}
\begin{equation}
 {a=a_0(t-t_0)^{\frac{1}{3}}.}
\end{equation}
Note that the potential energy density is negligible compared with the
kinetic energy density as $t\rightarrow\infty$ (indeed this asymptotic
solution may be obtained by simply neglecting $V$ in the field
equations). The curvature term in not negligible in this case so these 
results are only valid for $k=0$. 

If we include extra scalar fields but still assume that the potential
is dominated by a single exponential term $V_0
e^{\alpha\phi}e^{\beta\psi}$ then the situation just gets
worse because this potential must now dominate two kinetic terms to
yield inflation. One can make progress analytically by defining
$\Theta=\beta\phi-\alpha\psi$. Then the equations of motion for $\phi$
and $\psi$ imply that $\theta$ obeys
\begin{equation}
 {\ddot{\Theta}+3H\dot{\Theta}=0,}
\end{equation}
where the Hubble parameter is given by the Einstein constraint
equation with two scalar fields. This equation can be integrated to
give
\begin{equation}
 {\dot{\Theta}=A\exp\left(-3\int^t H(t')dt'\right),}
\end{equation}
where $A$ is a constant. If the scale factor grows sufficiently fast
then this term will be asymptotically negligible and
$\Theta\approx\textrm{constant}$ will be a good approximation. Then we
can write $\psi=\frac{\beta}{\alpha}\phi+\textrm{constant}$. Now
define
\begin{equation}
 {\theta=\frac{\sqrt{\alpha^2+\beta^2}}{\alpha}\phi}
\end{equation}
and the equations of motion reduce to those for a single scalar field
$\theta$ moving in an exponential potential with parameter
$\sqrt{\alpha^2+\beta^2}$. We can now apply the results derived above
to give the asymptotic behaviour of $\theta(t)$ and the scale
factor. The asymptotic behaviour of $\dot{\Theta}$ can then be found:
when $\alpha^2+\beta^2<6$, $\dot{\Theta}\propto
(t-t_0)^{-\frac{6}{\alpha^2+\beta^2}}$ and otherwise
$\dot{\Theta}\propto (t-t_0)^{-1}$. In both cases, $\dot{\phi}$ and
$\dot{\psi}$ are proportional to $(t-t_0)^{-1}$ so only when
$\alpha^2+\beta^2<6$ is it consistent to neglect $\dot{\Theta}$. In this
case one simply applies the single scalar result with parameter
$\sqrt{\alpha^2+\beta^2}$ to concluded that the solution is still
inflating at large times only when $\alpha^2+\beta^2<2$. By analogy
with the single scalar results one would only expect a significant
inflationary period from such solutions. 

If $\alpha^2+\beta^2>6$ then an asymptotic solution (for $k=0$) can
be found in analogy with the single scalar case by neglecting $V$. The
scalar equations of motion can be immediately integrated and the
result plugged into the Einstein constraint equation to give the scale
factor. The results are similar to the single scalar case. Our results
can obviously be generalised when there are more than two scalar
fields present. 

We can now return to our model obtained from eleven dimensional
supergravity. The final ($F$) term in $V$ is too steep to drive
inflation so we turn to the term coming from the (seven dimensional)
Ricci scalar. This depends upon the specific internal manifold that we
choose to squash but it is possible to extract the $\psi$ dependence
in the general case. To see this, note that the scalars $A_j$ all have
the same dependence on $\psi$ in \ref{eqn:inverse} hence the metric on
the internal space depends on $\psi$ only through the conformal factor
$e^{\frac{2\sqrt{14}}{21}\psi}$. It follows that the dependence
of the first term in $V$ on $\psi$ is given by a factor
$e^{-\frac{3\sqrt{14}}{7}\psi}$. This multiplies a $\phi_k$
dependent piece $\tilde{V}$. Since
$\left(\frac{3\sqrt{14}}{7}\right)^2>2$, the above work shows that it
is not possible to get inflationary behaviour driven by  a single
exponential term
in the potential. If inflationary solutions are possible then they
must arise from a combination of several such terms leading to a less
steep region of the potential, for example near a local
maximum. However the potential cannot possess a local maximum in
$\psi$ and only possesses a local minimum when $\tilde{V}<0$ and this
occurs at a negative value for $V$, which is obviously not suitable
for inflation. 

\section{Singular instantons}

The behaviour of the Hawking-Turok instanton \cite{ht1} corresponding to an
exponential potential can be analysed in a similar manner. The
instanton is assumed to possess an $O(4)$ symmetry so its metric can
be written
\begin{equation}
 {ds^2=d\sigma^2+b(\sigma)^2 d\Omega^2.}
\end{equation}
The Euclidean field equations are
\begin{equation}
 {\left(\frac{b'}{b}\right)^2=\frac{1}{3}\left(\frac{1}{2}\phi'^2-V\right)+\frac{1}{b^2},}
\end{equation}
\begin{equation}
 {b''=-\frac{1}{3}\left(\phi'^2+V\right)b,}
\end{equation}
\begin{equation}
 {\phi''+3\frac{b'}{b}\phi'=\frac{d V}{d\phi}.}
\end{equation}
Hawking and Turok consider solutions to these equations that are
regular at the North pole, where they look locally like four spheres,
and singular at the South pole. As the singularity is approached
they assume that the scalar kinetic term dominates its potential. We
shall investigate this assumption for $V=V_0e^{\alpha\phi}$. If the
curvature term $\frac{1}{b^2}$ is negligible in the Einstein
constraint equation then near the singularity we can write
\begin{equation}
 {\frac{b'}{b}=-\sqrt{\frac{1}{3}\left(\frac{1}{2}\phi'^2-V\right)}.}
\end{equation}
Substituting this into the scalar equation of motion and defining
$\Phi=\phi'e^{-\frac{1}{2}\alpha\phi}$ gives
\begin{equation}
 {\frac{d}{d\phi}\left(\frac{1}{2}\Phi^2\right)+\frac{\alpha}{2}\Phi^2-\sqrt{3\left(\frac{1}{2}\Phi^2-V_0\right)}\Phi-\alpha V_0=0.}
\end{equation}
Now let $\Phi=\sqrt{2V_0}\cosh x$, so we are assuming that the kinetic
term is larger than the potential term (otherwise the above
expressions do not make sense). For definiteness, take $x\geq 0$. 
This gives the equation
\begin{equation}
 {\frac{d x}{d\phi}=\frac{1}{2}\left(\sqrt{6}-\alpha\tanh x\right),}
\end{equation}
which can be integrated by defining $y=e^{2x}$ to give
\begin{equation}
 {e^{\frac{1}{2}\alpha(\phi-\phi_*)}=G(y)\equiv
 y^{\frac{\alpha}{2(\sqrt{6}+\alpha)}}\left|y+\frac{\sqrt{6}+\alpha}{\sqrt{6}-\alpha}\right|^{\frac{\alpha^2}{6-\alpha^2}}.}
\end{equation}
There are two cases to consider: \emph{i}) $\alpha<\sqrt{6}$ and
\emph{ii}) $\alpha>\sqrt{6}$. (Once again we can restrict $\alpha \geq
0$ through reversing the signs of $\alpha$ and $\phi$.) 

Case \emph{i}) $G(y)$ is a monotonically increasing function so $y$
becomes large as $\phi$ becomes large. Asymptotically we have
\begin{equation}
 {e^{\frac{1}{2}(\phi-\phi_*)}\approx
 y^{\frac{\alpha}{2(\sqrt{6}-\alpha)}}=e^{\frac{\alpha}{\sqrt{6}-\alpha}x} \Rightarrow x\approx\frac{1}{2}(\sqrt{6}-\alpha)(\phi-\phi_*).}
\end{equation}
This gives
\begin{equation}
 {\phi'e^{\frac{1}{2}\alpha\phi}=\Phi\approx\sqrt{2V_0}\cosh\frac{1}{2}(\sqrt{6}-\alpha)(\phi-\phi_*)\approx\sqrt{\frac{V_0}{2}}e^{\frac{1}{2}(\sqrt{6}-\alpha)(\phi-\phi_*)},}
\end{equation}
which is a differential equation that we can solve for $\phi$ to give
\begin{equation}
 {e^{-\frac{1}{2}\sqrt{6}\phi}\approx\sqrt{\frac{3V_0}{4}}e^{-\frac{1}{2}(\sqrt{6}-\alpha)\phi_*}(\sigma_f-\sigma),}
\end{equation}
where $\sigma_f$ is a constant of integration corresponding to the
coordinate of the singularity. The behaviour of the potential and
kinetic terms near the singularity is
\begin{equation}
 {V\propto (\sigma_f-\sigma)^{-\frac{2\alpha}{\sqrt{6}}},}
\end{equation}
\begin{equation}
 {\frac{1}{2}\phi'^2\approx \frac{1}{3}(\sigma_f-\sigma)^{-2},}
\end{equation}
so near the singularity the kinetic term will dominate. The behaviour
of the scale factor is easily obtained from the Einstein
constraint equation:
\begin{equation}
 {b\approx b_0(\sigma_f-\sigma)^{\frac{1}{3}}.}
\end{equation}
$b_0$ is a constant that is determined by matching the solution near
the South pole to the solution at the North pole. Note that it is
consistent to neglect the curvature term near the singularity. 
(These results agree with those obtained in \cite{copeland} 
but it was not pointed out there that they are only 
valid for $\alpha<\sqrt{6}$). 

For the singularity to be integrable, $\int d\sigma b^3V$ must
converge \cite{ht1} (one must include the boundary term $B_2$ to
derive this result). It is easy to see that it does in this case.

Case \emph{ii}) For $\alpha>\sqrt{6}$,
\begin{equation}
 {G(y)=y^{\frac{\alpha}{2(\alpha+\sqrt{6})}}\left|y-\frac{\alpha+\sqrt{6}}{\alpha-\sqrt{6}}\right|^{-\frac{\alpha^2}{\alpha^2-6}}.}
\end{equation}
Now $G(y)$ has a singularity at
$y=y_0\equiv\frac{\alpha+\sqrt{6}}{\alpha-\sqrt{6}}$ and tends to zero at large
$y$. Thus near the singularity the behaviour is
$\phi\rightarrow\infty$, $y\rightarrow y_0$ which implies
$x\rightarrow x_0$ and $\Phi\rightarrow\Phi_0$. This gives a
differential equation with solution
\begin{equation}
 {e^{\frac{1}{2}\alpha\phi}\approx\frac{2}{\alpha\Phi_0}(\sigma_f-\sigma)^{-1}.}
\end{equation}
Hence near the singularity the potential and kinetic terms behave as
follows
\begin{equation}
 {V\approx\frac{2}{\alpha^2\cosh^2 x_0}(\sigma_f-\sigma)^{-2}}
\end{equation}
\begin{equation}
 {\frac{1}{2}\phi'^2\approx\frac{2}{\alpha^2}(\sigma_f-\sigma)^{-2},}
\end{equation}
so now they only differ by a constant factor, which must be taken
account of in order to determine the solution for $b$. Substituting
these asymptotic solutions into the Einstein constraint equation and
eliminating $x_0$ in favour of $\alpha$ yields
\begin{equation}
 {b\approx b_0(\sigma_f-\sigma)^{\frac{2}{\alpha^2}}.}
\end{equation}
(Hence it is consistent to neglect the curvature term). So now we have
\begin{equation}
 {b^3 V\propto (\sigma_f-\sigma)^{\frac{6}{\alpha^2}-2},}
\end{equation}
but $\frac{6}{\alpha^2}-2<-1$ so the singularity is not integrable in
this case. 

In the two scalar case, arguments similar to those presented in the
previous section show that the singularity is only integrable for
$\alpha^2+\beta^2<6$, with obvious generalisation to more than two
fields.

In our model the potential $V$ contains a term coming from the
4-form. We shall assume that the correct analytic continuation of the
4-form to Euclidean signature is the one that leaves $V$
unchanged. This means that $F$ must be unchanged, so
$F_{\mu\nu\rho\sigma}$ must be imaginary in the Euclidean theory
(because $\sqrt{-g}\rightarrow i\sqrt{g}$) in agreement with the
discussion in \cite{bremer}. 

If the $F$-term is the dominant term in the potential near the
singularity then the above work shows that the singularity is not
integrable. Hence for a Hawking-Turok instanton to exist the dominant
part of the potential must come from the Ricci scalar of the internal
space. If one exponential term
$V_0e^{-\frac{3\sqrt{14}}{7}\psi}e^{\sum\lambda_i\phi_i}$ is dominant
then the condition for an integrable singularity is
$\sum\lambda_i^2<\frac{24}{7}$. This is not satisfied in the case of
the squashed $S^7$ considered above since
$\left(\frac{10\sqrt{21}}{21}\right)^2>\frac{24}{7}$. Hence the
squashed $S^7$ does not give an integrable singularity. For the
$S^1\times S^3\times S^3$ example we need at least one of the three
spheres to have negative curvature, so the dominant term must be one
of those in the second bracket in \ref{eqn:s3ricci}. Without loss of
generality we may assume it is the first one. If one writes this in
terms of the fields $\phi_k$ and $\psi$ then one finds that the
exponent is
$2\kappa\left(\sqrt{\frac{4}{5}}\phi_4+\sqrt{\frac{6}{5}}\phi_5+\sqrt{\frac{6}{7}}\phi_6+\frac{\sqrt{14}}{21}\psi\right)$,
so the sum of squares of $\phi_k$ coefficients is
$\frac{80}{7}>\frac{24}{7}$ hence the singularity is not integrable
in this case either.

\section{Singularity in eleven dimensions}

The examples provided by Garriga \cite{garriga2} are encouraging
evidence in favour of being able to obtain Hawking-Turok instantons in
four dimensions by dimensional reduction of higher dimensional
non-singular instantons. However Garriga's cosmological (as opposed to
flat) example is special because it requires the presence of a
cosmological constant in the higher dimensional theory. This always 
gives rise to a potential in the dimensionally reduced 
action that gives both inflation and an integrable singularity
on the instanton. Eleven dimensional supergravity does not have a
cosmological constant which is why we have been considering squashing
as an alternative mechanism of generating a positive potential in the
four dimensional effective action. Unfortunately it is easy to see
that the instantons of the type that we have considered remain
singular even when viewed from within the higher dimensional
framework. If one takes the trace of the eleven dimensional field
equations then one obtains
\begin{equation}
 {\hat{R}=\frac{\hat{\kappa}^2}{72}\hat{F}_{PQRS}\hat{F}^{PQRS},}
\end{equation}
which, upon substituting the solution for $\hat{F}_{PQRS}$ (and
remembering that a factor of $\sqrt{V_7}$ was absorbed into $F$),
becomes
\begin{equation}
 {\hat{R}=-\frac{\kappa^2}{3}F^2 e^{-\frac{2}{3}\sqrt{14}\kappa\psi}.}
\end{equation}
Since $\psi\rightarrow -\infty$ at the (four dimensional) singularity
independently of the sign of $R\left[M_7\right]$
(the exponents in $V$ are negative so $\psi\rightarrow -\infty$ rather
than $+\infty$ as considered above) one sees immediately that
$\hat{R}\rightarrow -\infty$ so the Hawking-Turok singularity is an
eleven dimensional curvature singularity.

\section{discussion and conclusions}

The results of the previous sections are rather disheartening: our
aim was to find a non-singular instanton in eleven dimensions that
gives rise to a Hawking-Turok instanton in four dimensions and an
inflationary period after continuing to Lorentzian signature. Instead
we have found that our model realizes neither of these
objectives. However the stumbling block appears to be the same in both
cases. Inflation was ruled out because the potential depends too
steeply on $\psi$. The singularity of the eleven dimensional instanton
is also due to the dependence on $\psi$. Note that the eleven
dimensional Ricci scalar is independent of $\phi_k$ so if some means
were found of fixing the size of the compactifying space (i.e. keeping
$\psi$ constant) then the instanton may become non-singular in eleven
dimensions even with $\phi_k\rightarrow\pm\infty$ (and hence
singular in four dimensions). The problem with keeping $\psi$ fixed
is in satisfying the eleven dimensional field equations. One would
have to introduce extra degrees of freedom in the metric of the
compactifying space, which involves going beyond squashing.

It is conceivable that by modifying our ansatz for the four-form more
interesting results might be obtained. The work in this paper is only
applicable to cosmological solutions using the Freund-Rubin ansatz
\cite{freund}. Bremer {\it et al} \cite{bremer} 
consider solutions with some more
general four-form configurations. However neither of their $S^7$
examples (round or squashed) appear to admit inflationary solutions
or instantons with integrable singularities. The Freund-Rubin ansatz
is attractive as an explanation of why there are four non-compact
spacetime dimensions but leads to a large negative
cosmological constant. In\cite{ht2} it was suggested that this may be
balanced by a contribution from supersymmetry breaking. Such symmetry
breaking would also have a dynamical effect at early times and might
generate corrections to the effective potential that make inflation
possible.

\appendix
\section{derivation of ricci tensor}

The non-zero Christoffel symbols for the metric \ref{eqn:ansatz} are given in 
terms of those for the ($D=11$) metric with $B=0$ by

\begin{eqnarray}
 \hat{\Gamma}^{\mu}_{\nu\rho} =
 \Gamma^{\mu}_{\nu\rho}+\delta^{\mu}_{\nu}B,_{\rho} +
 \delta^{\mu}_{\rho}B,_{\nu} - B,^{\mu}g_{\nu\rho} &\qquad&
 \hat{\Gamma}^{\mu}_{mn}=-e^{-2B}\sum_a {A_{a,}}^{\mu}e_m^a e_n^a
 \nonumber\\ \hat{\Gamma}^m_{n\rho} = \sum_a A_{a,\rho}e_a^m e_n^a &\qquad& \hat{\Gamma}^m_{np}=\Gamma^m_{np}.
\end{eqnarray}

Indices are raised with $g_{\mu\nu}$ and $e_m^a$ is the rescaled siebenbein.

The Ricci tensor is most easily computed using normal coordinates in the four
dimensional spacetime i.e. $\Gamma^{\mu}_{\nu\rho}=0$. (Note that we are not
free to choose normal coordinates on the whole eleven dimensional manifold
because such coordinates will not preserve the product form we have assumed
for the metric). The result is

\begin{eqnarray}
 \hat{R}_{\mu\nu} & = & R_{\mu\nu}-\left(\nabla^2B+\left(2B_{,\rho}+\sum_a
 A_{a,\rho}\right){B_,}^{\rho}\right)g_{\mu\nu}-\nabla_{\mu}\nabla_{\nu}\left(2B+\sum_a
 A_a\right) +{} \nonumber\\ & & {}+2B_{,\mu}B_{,\nu} +
 \sum_a\left(A_{a,\mu}B_{,\nu}+A_{a,\nu}B_{,\mu}\right) - \sum_a A_{a,\mu}A_{a,\nu},
\end{eqnarray}

\begin{eqnarray}
  \hat{R}_{mn} & = & R_{mn}[M_7] - e^{-2B} \sum_a e^a_m e^a_n
  \left[\nabla^2 A_a+{A_{a,}}^{\rho}\left(2B_{,\rho}+\sum_b A_{b,\rho}\right)\right].
\end{eqnarray}

$R_{mn}[M_7]$ is the Ricci tensor of the squashed internal manifold
computed treating the scalars $A_a$ as constants. Note that
substantial simplification occurs when we use the Einstein frame,
given by \ref{eqn:bsoln}.

\section{the squashed seven sphere}

As discussed in section III, $S^7$ can be regarded as the coset
$SO(5)/SO(3)$. The most general $SO(5)$ invariant metric on $S^7$
contains seven parameters. Here we restrict ourselves to a two
parameter subset. The particular squashing we use here is described in
more detail in \cite{duff}.

Using letters near the start of the Greek or Roman 
alphabet to denote tangent space indices, the
metric is given in terms of the siebenbein as $g_{mn}=\delta_{ab}e^a_me^b_n$
where
\begin{equation}
 \label{eqn:siebenbein}
 {e^0=d\mu,\qquad e^i=\frac{1}{2}\sin\mu\omega_i,\qquad e^{\hat{i}}=\frac{1}{2}e^{A(x)}(\nu_i+\cos\mu\omega_i).}
\end{equation}
The indices $i$ and $\hat{i}$ run from 1 to 3. $\mu$ is a coordinate taking
values in the range $[0,\pi]$. $A(x)$ is a scalar field that measures 
the amount of squashing. The round $S^7$ is given by $A=0$. 
The one forms $\nu_i$ and $\omega_i$ are given by
\begin{equation}
 \label{eqn:1forms}
 {\nu_i=\sigma_i+\Sigma_i,\qquad\omega_i=\sigma_i-\Sigma_i}
\end{equation}
where $\sigma_i$ and $\Sigma_i$ each satisfy the $SU(2)$ algebra:
\begin{equation}
 \label{eqn:algebra}
 {d\sigma_i=-\frac{1}{2}\epsilon_{ijk}\sigma_j\wedge\sigma_k,\qquad d\Sigma_i=-\frac{1}{2}\epsilon_{ijk}\Sigma_j\wedge\Sigma_k.}
\end{equation}
The tangent space components of the Ricci tensor are \cite{duff}
\begin{equation}
 {R_{ab} = \textrm{diag}(\alpha,\alpha,\alpha,\alpha,\beta,\beta,\beta),}
\end{equation}
where
\begin{equation}
 {\alpha = 3(1-\frac{1}{2}e^{2A}), \qquad \beta = e^{2A}+\frac{1}{2}e^{-2A}.}
\end{equation}

\medskip
\centerline {\bf Acknowledgements}
We have enjoyed useful discussions with Steven Gratton and Neil Turok.

\end{document}